\documentclass[aps,prl,twocolumn,superscriptaddress]{revtex4-1}

\usepackage{amsfonts}
\usepackage{amsmath}
\usepackage{amssymb}
\usepackage{graphicx}

\providecommand{\U}[1]{\protect\rule{.1in}{.1in}}

\begin{document}
\title{Electronic stopping power in gold: The role of $d$ electrons and the H/He anomaly}
\author{M. Ahsan Zeb}
\affiliation{Cavendish Laboratory, University of Cambridge,
Cambridge CB3 0HE, United Kingdom}

\author{J. Kohanoff}
\affiliation{Atomistic Simulation Centre, 
Queen's University, Belfast BT7 1NN, United Kingdom}

\author{D. S\'anchez-Portal}
\affiliation{Centro de F\'{\i}sica de Materiales CFM/MPC (CSIC-UPV/EHU), 
20018 San Sebasti\'an, Spain}
\affiliation{Donostia International Physics Center DIPC, 
Paseo Manuel de Lardiz\'abal 4, 20018 San Sebasti\'an, Spain}

\author{A. Arnau}
\author{J. I. Juaristi}
\affiliation{Centro de F\'{\i}sica de Materiales CFM/MPC (CSIC-UPV/EHU), 
20018 San Sebasti\'an, Spain}
\affiliation{Donostia International Physics Center DIPC, 
Paseo Manuel de Lardiz\'abal 4, 20018 San Sebasti\'an, Spain}
\affiliation{Departamento de F\'{\i}sica de Materiales, 
Facultad de Qu\'{\i}micas, UPV/EHU, 20018 San Sebasti\'an, Spain}

\author{Emilio Artacho}
\affiliation{Cavendish Laboratory, University of Cambridge,
Cambridge CB3 0HE, United Kingdom}
\affiliation{Nanogune and DIPC, 
Tolosa Hiribidea 76, 20018 San Sebasti\'an, Spain}
\affiliation{Basque Foundation for Science, Ikerbasque, 
48011 Bilbao, Spain}

\date{6 May 2012}

\begin{abstract}
  The electronic stopping power of H and He moving through gold 
is obtained to high accuracy using time-evolving density-functional 
theory, thereby bringing usual first-principles accuracies into this kind 
of strongly coupled, continuum non-adiabatic processes in condensed 
matter.
  The two key unexplained features of what observed experimentally 
have been reproduced and understood: 
  ($i$) The non-linear behaviour of stopping power versus velocity 
is a gradual crossover as excitations tail into the $d$-electron
spectrum;  and 
  ($ii$) the low-velocity H/He anomaly (the relative stopping powers 
are contrary to established theory) 
is explained by the substantial involvement of the $d$ electrons 
in the screening of the projectile even at the lowest velocities 
where the energy loss is generated by $s$-like electron-hole 
pair formation only.
\end{abstract}
\maketitle

  Non-adiabatic processes are at the heart of aspects of 
science and technology as important as radiation damage
of materials in the nuclear and space industries, and radiotherapy
in medicine.
  Yet, in spite of a long history, the quantitative understanding
of non-adiabatic processes in condensed matter and our ability to 
perform predictive theoretical simulations of processes coupling 
many adiabatic energy surfaces is very much behind what 
accomplished for adiabatic situations, for which first-principles 
calculations provide predictions of varied properties within a few 
percent accuracy.
  Substantial progress has been made for weakly non-adiabatic
problems such as the chemistry of vibrationally excited molecules 
landing on metal surfaces~\cite{Tully2009}, but not in the stronger
coupling regime of radiation damage.
  Recently, the electronic stopping power for swift ions in gold 
has been carefully characterized by experiments 
\cite{Figueroa2007,Markin2008,Markin2009}, showing flagrant 
discrepancies with the established paradigm for such problems 
\cite{Echenique1981,Review}, and only qualitative agreement with
time-dependent tight-binding studies \cite{Foulkes},
and with detailed studies for protons based on first 
principles \cite{Cantero2009}, leaving very fundamental questions 
unanswered in spite of the apparent simplicity of the system.
  Most notably the H/He anomaly: the present understanding 
predicts a stopping power for H higher than for He at 
low velocities \cite{Review}, which strongly contradicts 
the recent experiments \cite{Markin2009}.
  
  A particle moving through a solid material interacts 
with it and loses its kinetic energy to both the nuclei and the 
electrons inside it.
  At projectile velocities between 0.1 and 1 atomic units 
(a.u. henceforth) both the nuclear and the electronic contributions
to the stopping power (energy lost by the projectile 
per unit length) are sizeable~\cite{Foulkes}.
  Based on the jellium model (homogeneous electron gas)
the electronic stopping power, $S_e$, is predicted to be 
$S_e \propto v$ for a slow projectile traversing a metallic medium 
\cite{Ritchie1959,Kitagawa1974}.
  Such  behaviour has been observed experimentally 
in many $sp$-bonded metals \cite{Valdes1993,MartinezTamayo1996},
and the jellium model has allowed deep understanding of the
dynamic screening of the projectile and its relation to stopping
\cite{Quijada2007}.
  Even the jellium prediction of an oscillation of the proportionality 
coefficient with the projectile's atomic number $Z$ has been verified 
\cite{Review} and reproduced by ab initio atomistic simulations 
\cite{Hatcher2008}.
  However, phenomena that cannot be accounted for within the jellium 
paradigm have been described only qualitatively so far 
\cite{Valdes1994,Vargas1996, Valdes1997, Pruneda2007}.
  Experiments on noble metals Cu, Ag 
and Au, show pronounced nonlinearities in $S_e(v)$
\cite{Markin2008,Markin2009,Cantero2009,Blume1982,
Valdes1993,Valdes1994,Figueroa2007}.
  In the case of slow H and He ions in gold 
\cite{Markin2008,Markin2009,Figueroa2007}, $S_e(v)$
displays an increase in the slope roughly 
around $v\backsimeq0.18$ a.u.
  This is usually attributed to a threshold projectile velocity
needed to excite the $d$-band electrons that are relatively 
tightly bound.
  A model was developed based on the ab initio density of
electronic states and a stochastic treatment of 
excitations \cite{Valdes1997}, which reproduces the threshold 
for protons. 
  Here we obtain the non-linear $S_e(v)$ 
and the H/He anomaly with a general purpose 
ab initio method equally applicable to many other
radiation problems.

  We calculate the uptake of energy by the electrons in gold 
from a moving H or He ion in its $\langle100\rangle$ channel
by explicitly following the dynamics of the electrons coupled to the
projectile's motion, using time-evolving time-dependent density 
functional theory (TDDFT)  \cite{Runge1984}.
  We find very good quantitative agreement with some recent 
low energy ion scattering experiments on thin gold 
films \cite{Markin2008,Markin2009,Figueroa2007}.
  The results are analyzed in terms of the electronic 
excitations that are responsible for the energy loss, which
very clearly shows why the slope of $S_e$ 
increases with projectile velocity.
  In contrast to the usual idea that at low projectile velocity 
only electrons close to the Fermi energy contribute to the 
stopping, we find that there is a significant contribution 
from deep lying states even for a slow projectile.
  This means that at low velocities ($v< 0.2$ a.u.) the 
electrons accessible to excitations ($s$) are different from the 
ones involved in the screening of the projectile ($s+d$), the latter 
providing the excitation {\it mechanism}.

  We performed all calculations using the {\sc Siesta} method 
\cite{Ordejon1996} in its time evolving TDDFT implementation 
\cite{Tsolakidis2002}.
  We used the Perdew-Burke-Ernzerhof (PBE) version 
\cite{Perdew1996} of the Generalised Gradient Approximation 
(GGA) to the instantaneous exchange and correlation functional.
  Since we are only interested in the very low energy regime 
that is far below core electrons excitation thresholds, only valence 
electrons in Au are considered explicitly and a norm-conserving 
pseudopotential is used to describe the core electrons (up to the 
5$p$ sub-shell).
  Further details are found in \cite{SupplMat}.
  After placing the projectile in a [001] channel and finding the
DFT ground state, it is given an initial 
velocity along the $z$-direction while all gold atoms are initially 
quiescent.
  The system evolves by following Ehrenfest coupled electron-ion 
dynamics.
  On the time scale of the simulation ($\sim0.75-6.0$ fs for $v=0.05-0.50$ 
a.u.), the gold nuclei only gained negligible velocities and did not move 
significantly.
  We monitored the total energy of the electronic subsystem as a 
function of time.
  Once the transient related to the sudden start has 
disappeared \cite{SupplMat,Pruneda2007}, $S_e$ is extracted 
as the average rate of change of the electronic 
energy with the distance travelled by the projectile.

\begin{figure}[t]
\begin{center}
\includegraphics[width=0.5\textwidth]{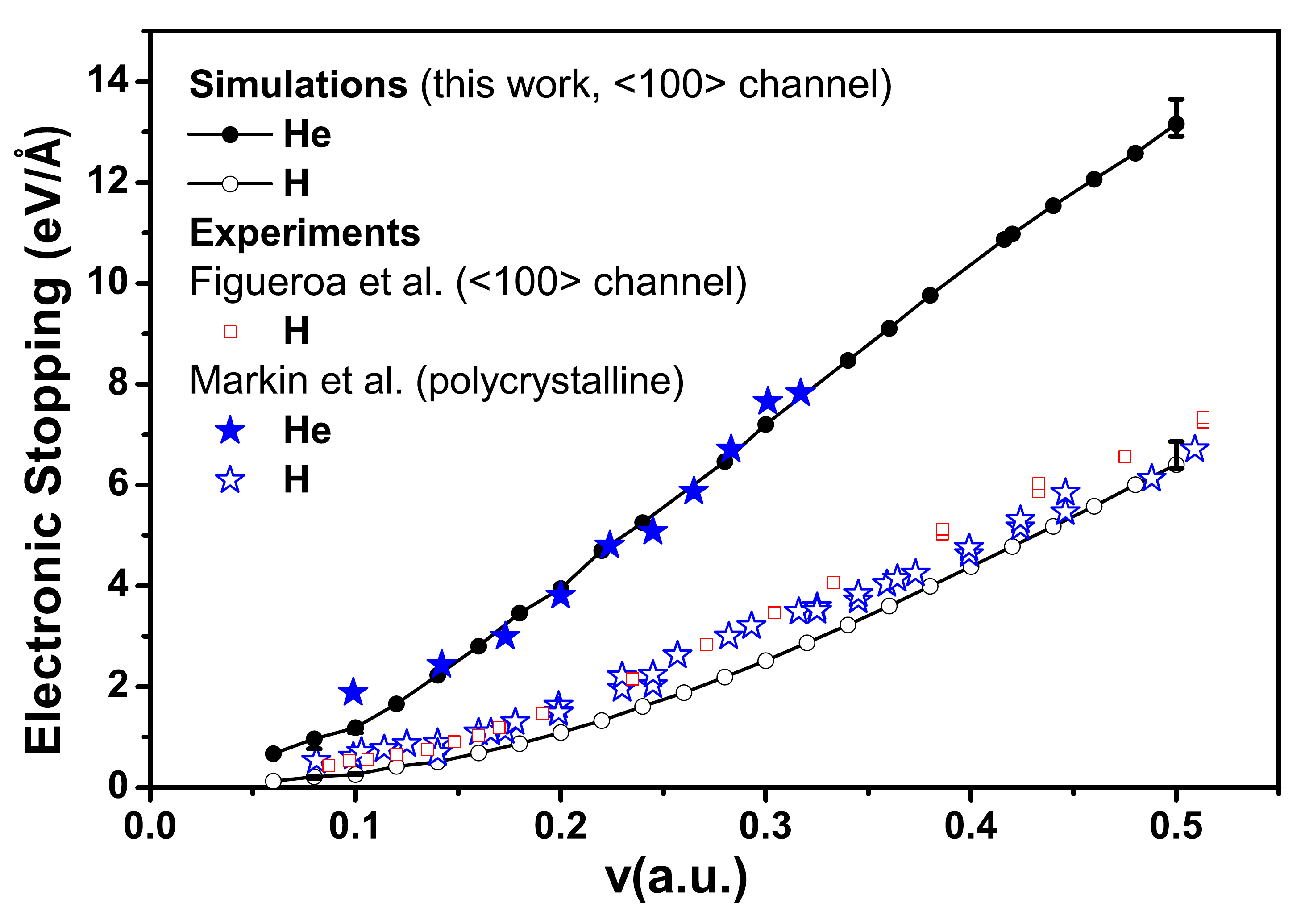}
\caption{\label{Stopping}{Electronic stopping power of H and 
He projectiles in gold as a function of projectile velocity. 
  Results of our simulations are compared with the experimental data 
from Refs.~\cite{Figueroa2007,Markin2008,Markin2009} on single and 
polycrystalline thin gold films.}}
\end{center}
\end{figure}

  Fig.~\ref{Stopping} shows our results for $S_e(v)$ for H and He 
projectiles in gold for the velocity range $v=0.06-0.50$ a.u.
  We also plot results of some recent
experiments performed on thin single crystal gold films oriented along
$\langle100\rangle$ \cite{Figueroa2007} and polycrystalline gold films 
\cite{Markin2008,Markin2009}.
  The agreement between our simulations and the experiments is 
noticeable.
  Although the stopping power is still underestimated (especially for 
H around $v = 0.3$ a.u.), no previous ab initio approach had this level of
agreement on the non-linear velocity dependence of the stopping 
power of real materials.
  Our results for the stopping power are well converged with respect 
to the basis size and the density of points on the real and the 
momentum space grids \cite{SupplMat}.
  A larger basis set for the projectile, however, (TZDP instead of DZP 
\cite{SupplMat}) increases the stopping power about 5\% at
$v=0.5$ a.u., but considerably less at low velocity.
  The error bars in Fig.~\ref{Stopping} indicate the dispersion in our 
results for $v = $0.08, 0.1 and 0.5 a.u. when the various parameters 
are varied, including the basis set \cite{SupplMat} (the bars for low
velocities are hardly larger than the size of the circles).
  The strict channelling in the simulation is partly behind the
observed underestimation: calculations for a 30\% smaller impact 
parameter give a 25\% increase in $S_e^{\rm H}$ at 
$v=0.28$ a.u. that reduces to 1\% at $v=0.5$ a.u.  

  We see a clear deviation from the linear behaviour around $v=0.2$ 
a.u. in $S_e$ of both H and He.
  This is unlike the $S_e \propto v$ of the uniform electron gas.
  It seems a plausible explanation that at low projectile velocity
only $s$-band electrons from the states around the Fermi energy 
contribute to the stopping and at higher velocity electrons in the 
$d$ band that lie relatively deeper in energy are also able 
to take part in it, resulting in an increase in the slope of $S_e$.
  Thus, comparisons have been made \cite{Figueroa2007,Markin2009} 
with jellium using the average electron density $n_e$
of the $s$ electrons ($r_{s}=3.01$ a.u., where 
$n_e^{-1}=\frac 4 3 \pi r_{s}^{3}$), using $r_s = 1.49$ a.u.,
corresponding to an effective number of $s$ and $d$ electrons
\cite{Markin2009}, or of the density in the $\langle100\rangle$ 
channel ($r_{s}=1.8$ a.u.). 
  However, the jellium predictions do not agree with the experimental 
results except at projectile velocities around $v=0.6$ a.u. in the latter 
case, despite the expectation that all the $d$-band electrons are 
active for a projectile velocity $v\geq0.47$ a.u. \cite{Valdes1994}.
  There is a further problem in the comparison with jellium: 
If we assume that at low velocity only
$s$ electrons are actively participating in the stopping mechanism, 
the jellium model predicts $S_e^{\rm H} > S_e^{\rm He}$
\cite{Echenique1981}, which is not the case.

\begin{figure}[t]
\begin{center}
\includegraphics[width=0.45\textwidth]{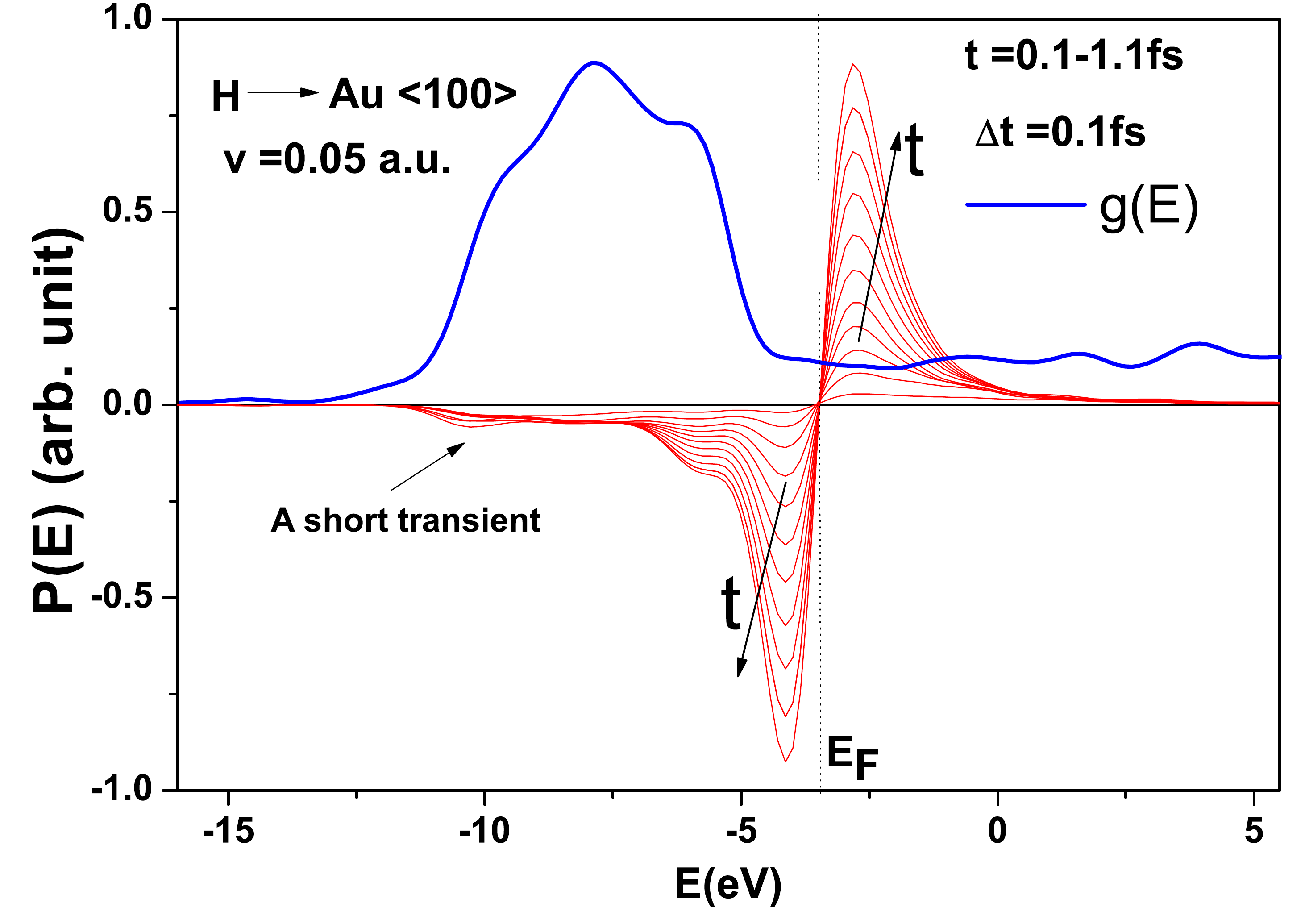}
\caption{\label{Projections}{
  (Color online) Excitation distribution $P(E)$ when H passes 
through gold with velocity 0.05 a.u., for time values between 
$t= 0.1$ fs and 1.1 fs in steps of $\Delta t = 0.1$ fs (light color; larger
amplitude for longer $t$; Gaussian broadening $\sigma=0.2$ eV).
  The dark curve is the electronic density of states $g(E)$
($\sigma=0.5$ eV). $P(E)$ and $g(E)$ are in different scales.}}
\end{center}
\end{figure}

  To explain the above inconsistencies and get a better idea of the energy 
loss mechanism we compute the changes in the electronic distribution 
due to the excitation of the electrons when a projectile propagates 
through the material.
Having $\{\left\vert \psi_{n}(t)\right\rangle \}$ and ${\mathbf X}(t)$, 
the set of evolved occupied KS states, and the corresponding 
atomic positions at time $t$,
  we calculate the adiabatic states $\{\left\vert \phi_{i},{\mathbf{X}}
\right\rangle \}$, i.e., the set of self-consistent static KS 
states for ${\mathbf X}(t)$.
  By projecting the evolved states onto 
the adiabatic states, $C_{in}=\left\langle 
\phi_{i},{\mathbf{X}(t)}|\psi_{n}(t)\right\rangle $, the 
density of occupied energy states $O(E)$ at time $t$ as a function
of energy $E$ are obtained as $O(E)=\sum_{i,n}{|C_{in}|^{2}
}\delta(E-E_{i})$.
  Here $E_{i}$ is the eigenvalue of the
adiabatic state $\left\vert \phi_{i}, {\mathbf{X}}\right\rangle $. 
  To compute the change in the electronic distribution or the
(electron-hole) excitation distribution, $P(E)$, we subtract the 
ground state electronic distribution from $O(E)$.
  That is, $P(E)=O(E)-\Theta(E_{F}-E)g(E)$, where $E_{F}$ is the 
Fermi energy of the system, $g(E)$ is the electronic density of 
states and $\Theta(E)$ is the Heaviside step function.

\begin{figure}[b]
\begin{center}
\includegraphics[width=0.41\textwidth]{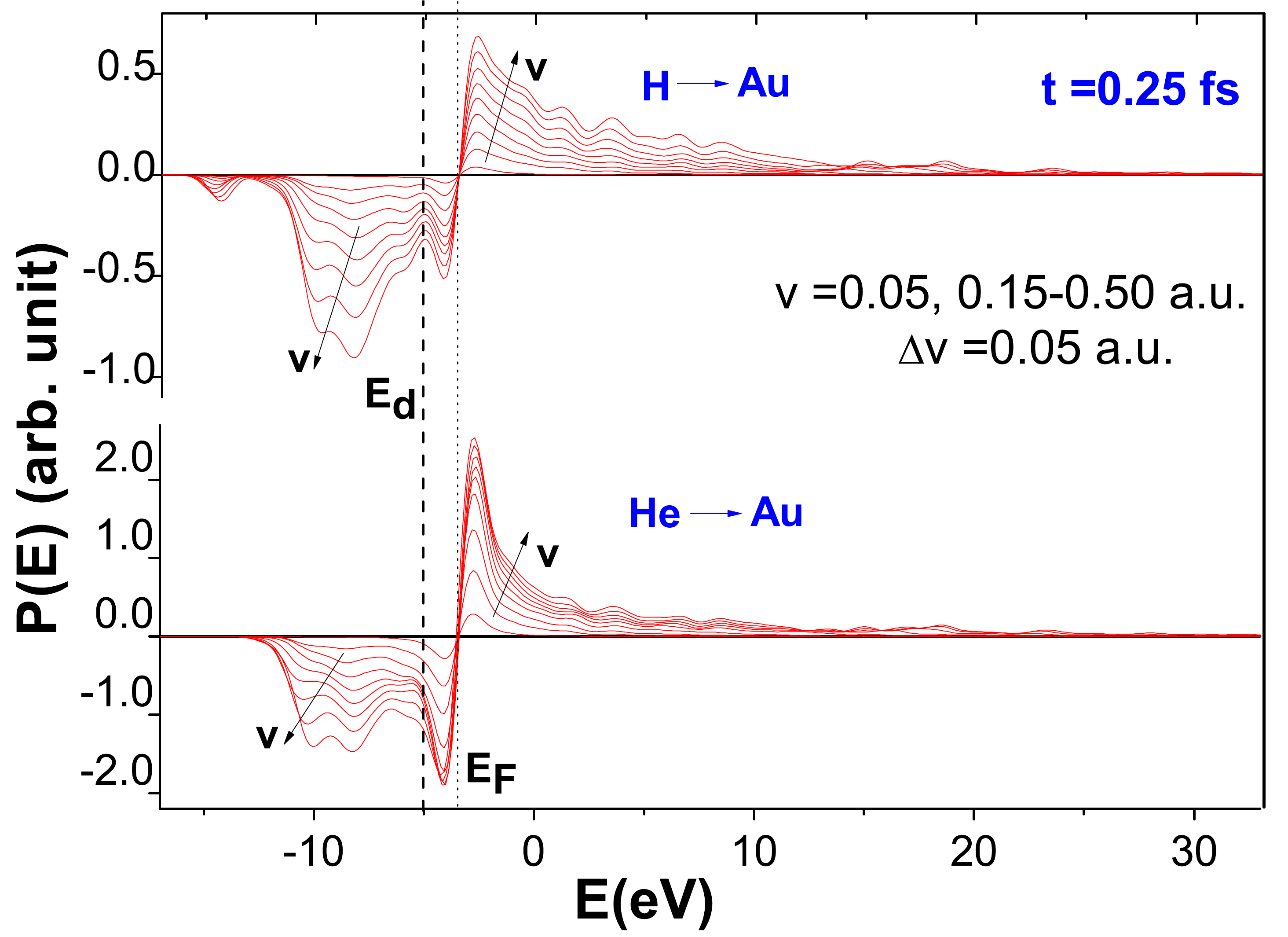}
\includegraphics[width=0.45\textwidth]{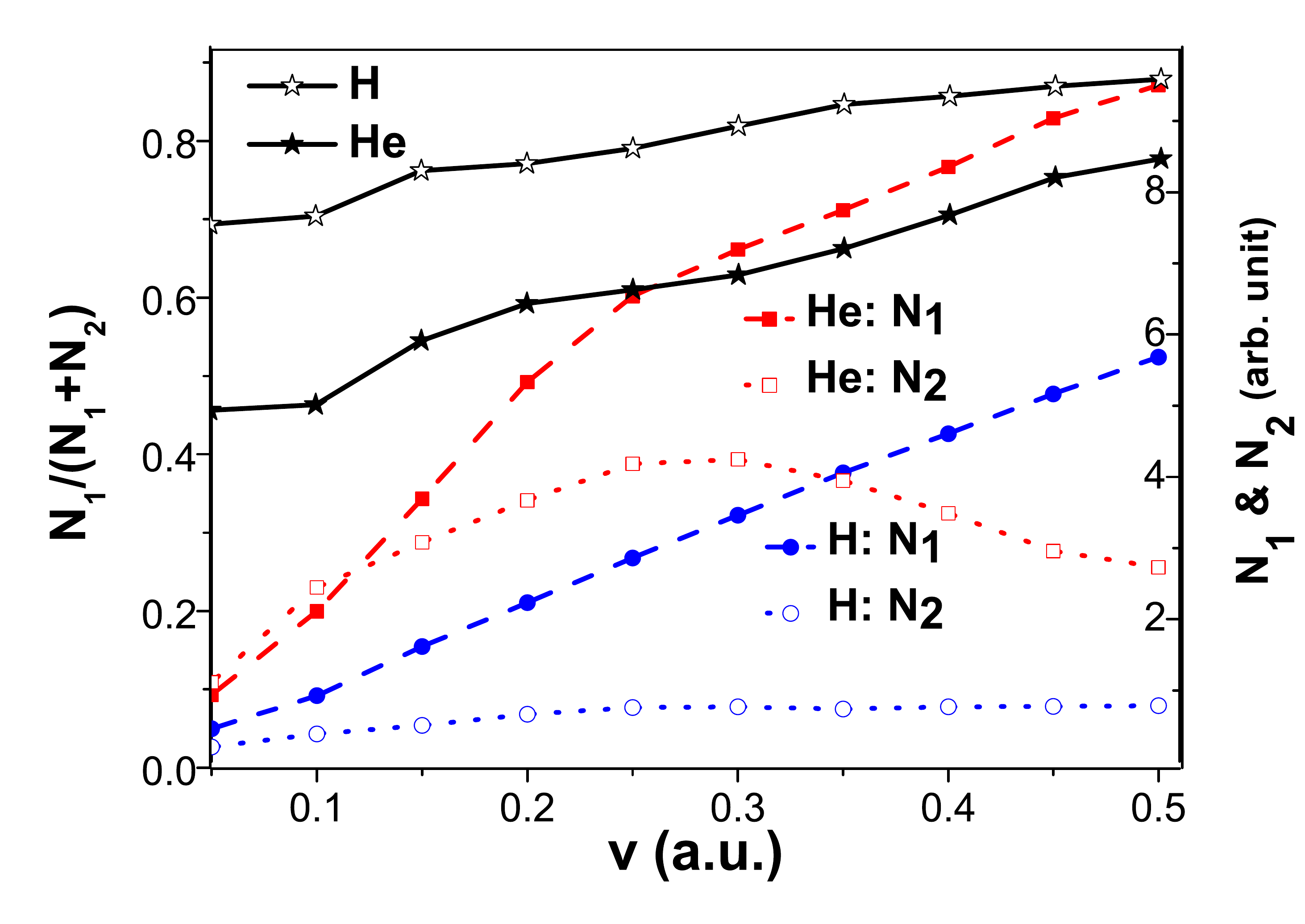}
\caption{\label{Weights}{
  Up: Excitation distribution $P(E)$ due to 
the passage of a H (top) or He projectile (bottom) in gold evaluated at 
$t = 0.25$ fs for various projectile velocities, $v = 0.05-0.50$ a.u. in 
steps of 0.05 a.u.
  Increased projectile velocity gives curve with larger amplitude
(indicated by arrows).
  The dashed and dotted vertical lines show the upper edge of the 
gold's $5d$-band $E_{d}$ and the Fermi energy $E_{F}$.
  Down:  Number of empty states below and above 
$E_{d}$, $N_{1}$ and $N_{2}$, and fraction $N_{1}/(N_{1}+N_{2})$ 
versus projectile velocity, due to the excitations for H or He.}}
\end{center}
\end{figure}

  Fig.~\ref{Projections} shows the excitation distribution $P(E)$ as a 
function of energy at various instants from $t=0.1$ fs to $t=1.1$ fs 
with an interval $\Delta t=0.1$ fs, for the passage of a H atom 
in gold along $\langle100\rangle$ with velocity $v=0.05$ a.u.
  The electronic density of states of the bulk Au host $g(E)$ is also 
plotted in Fig.~\ref{Projections}.
  The negative and the positive values of $P(E)$ show the density 
of empty and filled states below and above $E_F$, 
respectively, due to the electronic excitations caused by the 
moving projectile.
  Notice that
despite being very slow ($v=0.05$ a.u), the projectile is able to 
excite relatively tightly bound $d$-band electrons.
  A short initial transient behaviour is noticeable in 
Fig.~\ref{Projections}:
at energies deep below the Fermi energy, the number of empty 
states becomes larger initially, requiring a short time to adjust to a 
stationary regime.
  This is because in our simulations the projectile is a static impurity 
atom at $t=0$ that suddenly acquires a finite velocity resulting in a
large initial perturbation. 

  To see how the excitation distribution after the transient depends 
on the velocity of the projectile, we plot $P(E)$ against $E$ in 
Fig.~\ref{Weights} at $t=0.25$ fs for various projectile velocities,
between 0.05 a.u. and 0.50 a.u.
  We see that, compared to the states just below the Fermi energy, 
the number of excitations from deep inside the $d$-band increases 
more quickly with the velocity of the projectile.
  This means that the effective number of $d$-band electrons involved 
directly in excitations provoking the stopping process increases with the 
projectile velocity.
  To see this more clearly, we separated the energy window into two 
parts at the upper edge of the $d$-band at energy $E_{d}$ and calculated 
the total number of excitations $N_{1}$ and $N_{2}$ from the states below 
and above $E_{d}$ for a constant distance travelled by the projectile.
  We find that $P(E)\propto t$ after the initial transient so we can estimate
$N_{1}$ and $N_{2}$ as $N_{1}=\frac{1}{v}\int_{-\infty}^{E_{d}}|P(E)|dE$ 
and $N_{2}=\frac{1}{v}\int_{E_{d}}^{E_{F}}|P(E)|dE$, which we did using 
$P(E)$ at $t=0.25$ fs.
  In the inset of Fig.~\ref{Weights} we plot $N_{1}$ and $N_{2}$ and the
fraction $N_{1}/(N_{1}+N_{2})=N_{1}/N$ against the projectile velocity as
dashed, dotted and solid lines for H and He projectiles.
  We see that $N_{1}$ and $N_{1}/N$ increase with $v$ for both 
projectiles. 
  For H, $N_{2}$ increases and saturates whereas for He it 
increases up to $v=0.3$ a.u. but decreases for a faster projectile.
  Since there is one $s$ electron and ten $d$ electrons and 
$N_{1}$ also includes the contribution from the s-band states, ideally 
it should tend to $N_{1}/N \sim 10/11=0.909$ for high projectile 
velocity.
  $N_{1}/N$ reaches only $0.88$ and $0.78$ for H and He at 
$v=0.5$ a.u.
  Although the fraction of excitations from the deep lying states is higher 
for H, the absolute number is lower, as can be seen in the figure.
  Furthermore, in the case of H, $N_{1}>N_{2}$ for the whole velocity range 
shown whereas for He, $N_{2}>N_{1}$ in the very low velocity range. 

\begin{figure}[t]
\begin{center}
\includegraphics[width=0.38\textwidth]{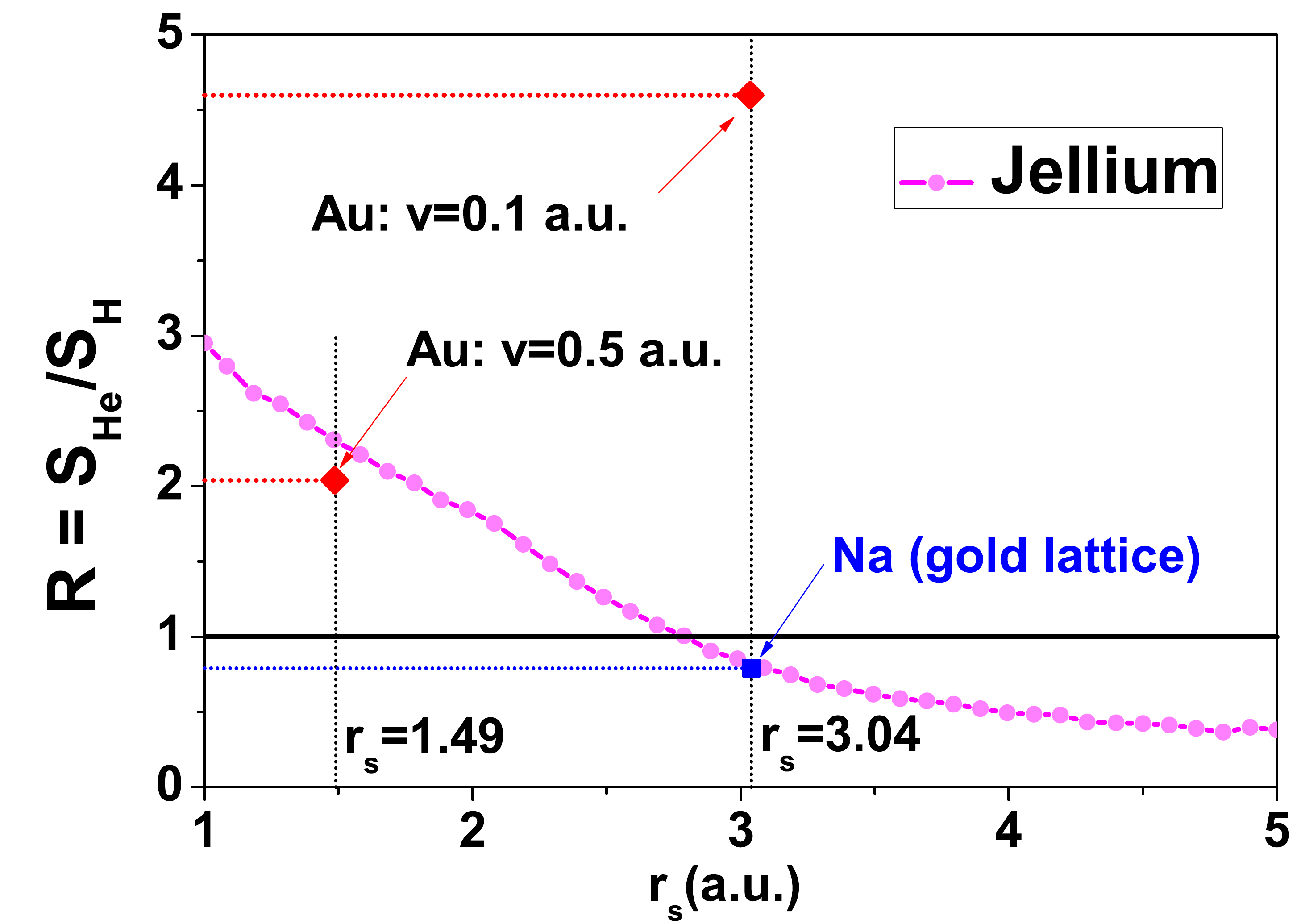}
\includegraphics[width=0.35\textwidth]{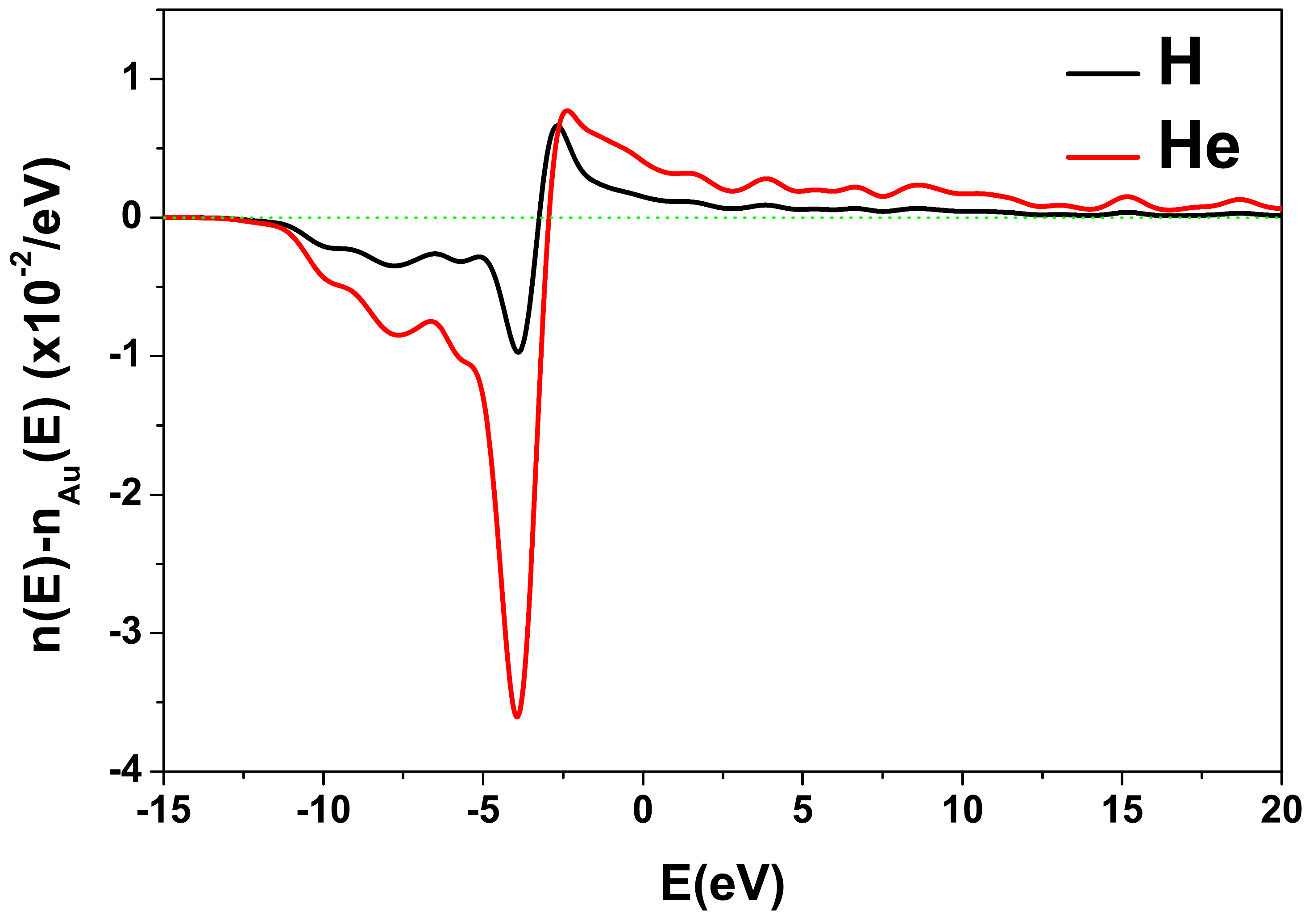}
\caption{\label{HvsHe} {
  Up: The curve shows $R=S_e^{\rm He} / S_e^{\rm H}$ for
jellium versus the electron density parameter $r_{s}$ 
\cite{Echenique1981}.
  The values of $R$ obtained for Au for $v=0.1$ and $0.5$ a.u. 
are associated with $r_{s}=3.04$ and $1.49$ a.u, respectively, 
following Ref.~\cite{Markin2009}.
  The calculated ratio $R$ for a system made of Na atoms in 
bulk Au positions, $r_{s}=3.04$, is also presented.
  Down: Projection of KS states for the system with projectile onto
the KS states of bulk Au, for H and He, subtracting the
Au density of states.}}
\end{center}
\end{figure}

  We address now the low-velocity H/He anomaly. 
  Fig.~\ref{HvsHe} presents the ratio $R=S_e^{\rm He} / S_e^{\rm H}$ 
in jellium~\cite{Echenique1981}.
   The values of $R$ for Au for  $v=0.1$ and $v=0.5$ a.u. 
are plotted on the two dotted vertical lines at $r_{s}=3.04$ 
and $r_{s}=1.49$ a.u., which correspond to 1 and 8.24 
electrons per bulk unit cell, i.e., the $s$ electrons and the 
effective number of valence electrons ($s$ and $d$)
that fit the plasmon pole for bulk Au~\cite{Markin2009}.
  We see that for the faster projectile $R$ is close to the jellium 
value and significantly larger than 1.
  However, for the slower one, we obtain $R=4.7$,
in clear disagreement with the jellium value of 0.79, but in
agreement with experiment.
  We also plot $R$ for the fictitious system
built by putting Na atoms in the Au positions, which 
corresponds to an electron gas with $r_{s}=3.04$.
  The plot shows a perfect agreement for the $R$ values of 
Na and jellium \cite{footnotevalues}.
  These differences between jellium (or Na) and gold are thus
due to the presence of gold's $d$ electrons.
  This is consistent with the fact that, even if a slow projectile 
were unable to excite the $d$-band electrons appreciably, 
the presence of the projectile in gold constitutes a large 
static perturbation for the $d$ electrons.
  This can be clearly seen by calculating the projection of the
ground state of the gold with the projectile onto that without it and
obtaining a distribution analogous to $P(E)$, now
describing the static screening of the projectile. i.e.,
projecting the wave-functions of Au with the projectile onto
the states of pure Au (Fig.~\ref{HvsHe}). 
  This means that for a slow projectile the response 
of the electrons in gold is far from the one described by the 
homogeneous electron gas model that includes just the 
$s$-band electrons.


  To summarize, we have shown that realistic non-adiabatic 
stopping of projectiles in real metals can now be described from 
first-principles with acceptable accuracy, even at the 
Ehrenfest dynamics level.
  We used it to calculate the electronic energy loss on passage 
of H and He through Au and find good quantitative 
agreement with experiments.
  Many problems can now be addressed with this technique in the
fields mentioned in the introduction.

  JK thanks the Wellcome Trust for a Flexible Travel award. 
  MAZ acknowledges support of the Islamic Development Bank.
  DSP, JIJ and AA acknowledge support of 
the University of the Basque Country UPV/EHU, under Grant IT-366-07,
and of the Spanish Ministerio de Ciencia e Innovaci\'on under Grant 
FIS2010-19609-C02-00. The calculations were done using the CamGrid 
high-throughput facility of the University of Cambridge.


\begin{thebibliography}{99}

\bibitem{Tully2009} 
N. Shenvi, S. Roy and J. C. Tully, and refs. therein, 
Science {\bf 326}, 829 (2009).

\bibitem {Figueroa2007}
E. A. Figueroa, E. D. Cantero, J. C. Eckardt, 
G. H. Lantschner, J. E. Vald\'es, and N. R. Arista, 
Phys. Rev. A {\bf 75}, 010901 (2007).

\bibitem {Markin2008}
S. N. Markin, D. Primetzhofer, S. Prusa, M. Brunmayr, 
G. Kowarik, F. Aumayr, and P. Bauer, 
Phys. Rev. B {\bf 78},195122 (2008).

\bibitem {Markin2009}
S. N. Markin,D. Primetzhofer, M. Spitz, and P. Bauer,
 Phys. Rev. B {\bf 80}, 205105 (2009).

\bibitem {Echenique1981}
P. M. Echenique, R. M. Nieminen and R. H. Ritchie, 
Solid State Comm. {\bf 37},779 (1981).

\bibitem{Review} 
P. M. Echenique, F. Flores and R.H. Ritchie, 
Solid State Physics {\bf 43}, eds. H. Ehrenreich and 
D. Turnbull, Academic Press, p. 230 (1990).

\bibitem{Foulkes} C. P. Race, D. R. Mason, M. W. Finnis, 
W. M. C. Foulkes, A. P. Horsfield, and A. P. Sutton,
Rep. Prog. Phys. {\bf 73}, 116501 (2010), 
and refs. therein.

\bibitem {Cantero2009}
E. D. Cantero, G. H. Lantschner, J. C. Eckardt, 
and N. R. Arista, Phys. Rev. A {\bf 80}, 032904 (2009).

\bibitem {Ritchie1959}
R. H. Ritchie, Phys. Rev. {\bf 114}, 644 (1959).

\bibitem {Kitagawa1974}
M. Kitagawa and Y. H. Ohtsuki, 
Phys. Rev. B {\bf 9}, 4719 (1974).

\bibitem {Valdes1993}J. E. Vald\'es and G. Mart\'{\i}nez-Tamayo, 
G .H. Lantschner, J. C. Eckardt and N. R. Arista, 
Nucl. Instrum. Meth. B {\bf 73}, 313 (1993).

\bibitem {MartinezTamayo1996}
G. Mart\'{\i}nez-Tamayo, J. C. Eckardt, G. H.
Lantschner, and N. R. Arista, Phys. Rev. A {\bf 54}, 3131 (1996).

\bibitem{Quijada2007} M. Quijada, A. G. Borisov, I. Nagy, 
R. D\'{\i}ez Mui\~no, and P. M. Echenique, 
Phys. Rev. A {\bf 75}, 042902 (2007).

\bibitem{Hatcher2008} 
R. Hatcher, M. Beck, A. Tackett, and S. T. Pantelides,
Phys. Rev. Lett. {\bf 100}, 103201 (2008).

\bibitem {Valdes1994}
J. E. Vald\'es, J. C. Eckardt, G. H. Lantschner, and
N. R. Arista, Phys. Rev. A {\bf 49}, 1083 (1994).

\bibitem {Vargas1996}
P. Vargas, J. E. Vald\'es and N. R. Arista, 
Phys. Rev. A {\bf 53}, 1638 (1996).

\bibitem {Valdes1997}
J. E. Vald\'es , P. Vargas and N. R. Arista, 
Phys. Rev. A {\bf 56}, 4781 (1997).

\bibitem {Pruneda2007} J. M. Pruneda {\it et al.} 
Phys. Rev. Lett. {\bf 99}, 235501 (2007).

\bibitem {Blume1982}
R. Blume, W. Eckstein, and H. Verbeek, 
Nucl. Instrum. Methods {\bf 168}, 57 (1980); {\bf 194}, 67 (1982).

\bibitem {Runge1984}E. Runge and E. K. U. Gross, 
Phys. Rev. Lett. {\bf 52}, 997 (1984).

\bibitem {Ordejon1996}
P. Ordej\'on, E. Artacho, and J. M. Soler, 
Phys. Rev. B {\bf 53},10441 (1996); 
J. M. Soler {\it et al.}, 
J. Phys. Condens. Matter {\bf 14}, 2745 (2002).

\bibitem {Tsolakidis2002}
A. Tsolakidis, D. S\'anchez-Portal, and R. M. Martin, 
Phys. Rev. B {\bf 66}, 235416 (2002).

\bibitem {Perdew1996} 
J. P. Perdew, K. Burke and M. Ernzerhof, 
Phys. Rev. Lett. {\bf 77}, 3865 (1996).

\bibitem{SupplMat}
See details in supplementary materials.


\bibitem{footnotevalues} 
The stopping power values for H and He in this Na are 
3.902 eV/\AA\ and 3.083 eV/\AA, respectively, for $v=0.5$ a.u., 
and 1.024 eV/\AA\ and 0.715  eV/\AA, for $v=0.1$ a.u.; 
the corresponding values for jellium as extracted from 
Ref. \cite{Echenique1981} are
7.741 eV/\AA\ and 17.874 eV/\AA, respectively, for $v=0.5$ a.u., 
and $r_s=1.49$ Bohr; 
and 0.861 eV/\AA\ and 0.710 eV/\AA, for $v=0.1$ a.u.
and $r_s=3.04$ Bohr. 


\end{thebibliography}
\end{document}